\documentclass[aps,pre,epsfig,floats,superscriptaddress,amssymb,amsmath,floatfix,preprint]{revtex4-1}
\usepackage{graphicx}

\begin{document}

\def\eq{\begin{eqnarray}}
\def\qe{\end{eqnarray}}
\def\bG{{\bf G}}
\def\bI{{\bf I}}
\def\bR{{\bf R}}
\def\bS{{\bf S}}
\def\be{{\bf e}}
\def\bg{{\bf g}}
\def\bn{{\bf n}}
\def\br{{\bf r}}
\def\bt{{\bf t}}
\def\bv{{\bf v}}
\def\bzeta{\mbox{\boldmath $\zeta$}}
\def\brho{\mbox{\boldmath $\rho$}} 
\def\simle{\;\lower3pt\hbox{$\stackrel{\textstyle <}{\sim}$}\;}

\title{
Hydrodynamic Synchronization
between Objects with Cyclic Rigid Trajectories
}
\author{Nariya Uchida}
\email{uchida@cmpt.phys.tohoku.ac.jp}
\affiliation{Department of Physics, Tohoku University, Sendai, 980-8578, Japan}

\author{Ramin Golestanian}
\email[]{ramin.golestanian@physics.ox.ac.uk}
\affiliation{Rudolf Peierls Centre for Theoretical Physics, University of Oxford, Oxford OX1 3NP, UK}

\date{\today}

\begin{abstract}
Synchronization induced by long-range hydrodynamic interactions is
attracting attention as a candidate mechanism behind coordinated
beating of cilia and flagella.
Here we consider a minimal model of hydrodynamic synchronization
in the low Reynolds number limit. The model consists of rotors,
each of which assumed to be a rigid bead making a fixed trajectory
under periodically varying driving force.
By a linear analysis, we derive the necessary and sufficient
conditions for a pair of rotors to synchronize in phase.
We also derive a non-linear evolution equation for their
phase difference,
which is reduced to minimization of an effective potential.
The effective potential is calculated for a variety of trajectory
shapes and geometries (either bulk or substrated),
for which the stable and metastable states of
the system are identified.
Finite size of the trajectory induces asymmetry of
the potential, which also depends sensitively
on the tilt of the trajectory.
Our results show that flexibility of cilia or flagella
is {\it not} a requisite for their synchronized motion,
in contrast to previous expectations.
We discuss the possibility to directly implement the model
and verify our results by optically driven colloids.
\end{abstract}


\maketitle

\section{Introduction}  \label{sec:intro}

Coordinated cyclic beating of elastic organelles such as cilia and eukaryotic flagella
serve a multitude of functions in living organisms, ranging from motility to fluid
transport and polarity symmetry breaking in developing embryos \cite{Gray,bray,hamada}.
It has long been known that the beating cycle of cilia has a characteristic
asymmetry, with two distinct parts described as power stroke and recovery
stroke \cite{Blake-Sleigh}, and that the cyclic pattern could lead to
metachronal waves (of varying kinds) \cite{Knight-Jones} in dense arrays of cilia \cite{Blake72,GL97,GL99}. While the necessity of this asymmetry for generating symmetry
breaking fluid flow or propulsion could be easily understood from the time-reversal
symmetry properties of the Stokes equation for viscous fluid flow, what exactly
constitute the minimal conditions for synchronization and coordination between
two or more of such cyclically beating organelles is a subject of current
investigation \cite{GYU2011}.

There have been a number of systematic experimental studies in a variety of systems
to probe whether viscous hydrodynamic interaction alone can lead to synchronization as
Taylor \cite{Taylor} originally proposed. The experiments, which all verify
the existence of hydrodynamic synchronization, range from studying macroscopic model
flagella in highly viscous silicone oil (such that the low Reynolds number condition
was maintained) \cite{Kim-PNAS-03,QJ09} to probing the relative phase dynamics in pairs
of beating eukaryotic flagella \cite{Goldstein-Science-09,GP09}, to tracking colloidal
linear oscillators using optical tweezers equipped with feedback control \cite{Kotar-10}
and light driven asymmetrically micro-fabricated rotors \cite{DiLeonardo-1}. Experiments
on carpets of bacteria with active flagella \cite{Berg} and arrays of artificial
magnetically actuated cilia \cite{Vilfan-10,Shields-10,Bartolo-1} have revealed
collective effects mediated by hydrodynamic interactions, such as complex flow
patterns and collective phase shifts.

Theoretical studies of metachronal coordination and synchronization of cyclically
beating organelles have been performed using models and descriptions of varying
levels of complexity, ranging from simple models of coupled oscillators to actuated
beads and more elaborate elastic filament models \cite{Blake72,GL97,GL99,LB02,LJ68,KP04,RS05,KN06,VJ06,RL06,GJ07,NEL08,EL09,UG10a,UG10b,UG11,OV11}.
While the more realistic beating elastic filament descriptions are crucial
for understanding detailed properties of metachronal waves, the simpler
actuated bead models that typically have a minimal number of degrees of freedom
could be useful in understanding what key ingredients are needed for hydrodynamic
synchronization to occur, and under what conditions such dynamical states could be stable.
In the course the studies of actuated bead models, one of the questions that
have been discussed in whether or not beads following rigid trajectories could
lead to synchronization. The discussion started when it was shown that
two rigid helices that are rotating under constant torque cannot synchronize \cite{KP04}
while with an added a small flexibility, say to the axis of rotation, the system
can synchronize \cite{RS05}. Further studies of the actuated bead model followed
the prescription of always having a flexible element, and somehow this was later on
erroneously interpreted by many authors as a necessary condition for synchronization.
In an earlier publication  \cite{UG11}, we showed that flexibility is not a necessity,
and that beads following rigid trajectories could lead to synchronization provided
the shape of the trajectories and the beating force profile satisfy certain conditions.
The aim of this paper is to present a thorough discussion of how synchronization
could be achieved for rigid trajectories in a variety of cases.

The rest of the paper is organized as follows.
In Section II, we introduce the model and derive the coupled-oscillator
equation. In Section III,
generic conditions for synchronization
are derived by linear stability analysis,
and then applied to some specific trajectories and force profiles.
In Section IV, we discuss flow properties, especially
the net flow and energy dissipation rate in the synchronized state.
In Section V, we describe the nonlinear time-evolution equation
for the phase difference by an effective potential, which is then
used to determine the stable and metastable stationary states
for various trajectories and force profiles.
In Section VI, the effect of flexibility is taken into account
in a model of beads driven by moving harmonic traps.
Finally in Section VII, before conclusion, we discuss the implications
of our results to biological systems and their direct verification
by optically driven colloids.

\section{Model}
\subsection{Dynamical Equations}

\begin{figure}[t]
\includegraphics[width=0.79\columnwidth]{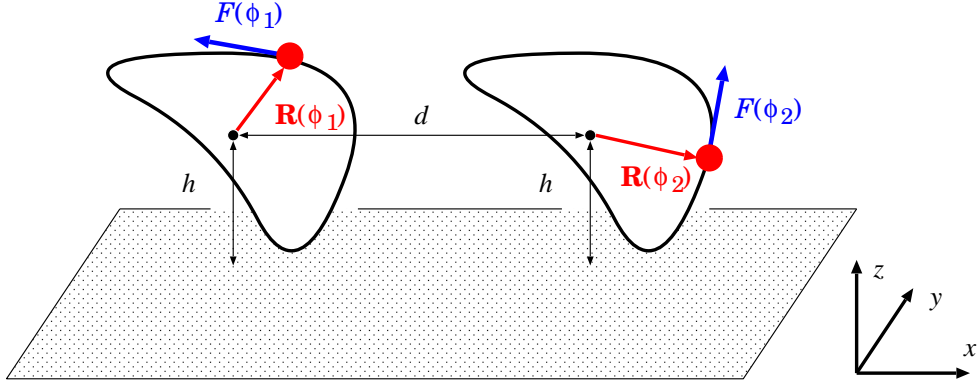}
\caption{
A pair of rotors with the trajectory shape
specified by ${\bf R}(\phi_i)$ ($i=1,2$).
Each bead is driven by the tangential force $F(\phi_i)$.
The centers of the trajectories are both on the $x$-axis.
\label{fig:twobeads}
}
\end{figure}

We consider a pair of rotors (indexed by $i=1,2$)
and assume that each is a spherical bead of radius $a$
that follows a fixed periodic trajectory
${\bf r}_i = {\bf r}_i(\phi_i)$, where $\phi_i = \phi_i(t)$ is
the phase variable
with the period $2\pi$ [see Fig. \ref{fig:twobeads}].
The bead is driven by an active force $F_i = F_i(\phi_i)$
that is tangential to the orbit and is an arbitrary function of the phase.
We assume that the two rotors are situated in parallel to each other, and
that the center points of the trajectories are
at height $h$ from a flat substrate.
The $xy$-plane is taken along the substrate,
with the $x$-axis parallel to the line connecting the center points,
and the $z$-coordinate is taken vertically to the substrate.

The hydrodynamic drag force acting on the $i$-th bead is
written in the form $\bg_i = \bzeta \cdot [\bv(\br_i) - \dot{\br}_i]$.
The friction coefficient tensor $\bzeta$ depends on the height $z$
of the bead from the substrate.
However, the dependence is $O(a/z)$ and we neglect it
by assuming $a \ll z$. Then the friction coefficient tensor
is expressed by the friction coefficient $\zeta_0 = 6\pi \eta a$
as $\bzeta = \zeta_0 \bI$.
The tangential component of the drag force
is balanced by the driving force acting on each rotor, namely,
\begin{math}
F_i + {\bf t}_i \cdot {\bf g}_i = 0,
\end{math}
where ${\bf t}_i$ is the tangential unit vector of the orbit given by
${\bf t}_i = {{\bf r}'_i}/{|{\bf r}'_i|}$ with ${\bf r}'_i = {d{\bf r}_i}/{d\phi_i}$.
Substituting the expression for the drag force with
$\dot{\bf r}_i =  {\bf r}'_i \dot\phi_i$ into the force balance equation, we
obtain the phase velocity as
\begin{math}
\dot\phi_i
= \omega_i
+ {{\bf t}_i \cdot {\bf v}({\bf r}_i)}/{|{\bf r}'_i|},
\end{math}
where
\begin{math}
\omega_i(\phi_i)
= {F_i(\phi_i)}/{\zeta_0 |{\bf r}_i'|}
\end{math}
is the intrinsic phase velocity.
The reaction force $- {\bf g}_i$ exerted by the bead on the fluid
generates the flow field
\begin{eqnarray}
{\bf v}({\bf r}) = - \sum_j {\bf G}({\bf r}, {\bf r}_j) \cdot {\bf g}_j
\simeq
\sum_{j} \zeta_0{\bf G}({\bf r}, {\bf r}_j) \cdot {\bf r}'_j \omega_j.
\label{vflow}
\end{eqnarray}
Here, ${\bf G}({\bf r}, {\bf r}_j)$ is the Green function
of the Stokes equation with the no-slip boundary condition
at the substrate (Blake tensor).
We will give its expression in the next subsection.
On the RHS of Eq. (\ref{vflow}), we assumed $|{\bf r} - {\bf r}_j| \gg a$
and retained the leading order term
with respect to
$\zeta_0{\bf G}({\bf r}, {\bf r}_j) = O(a/|{\bf r}-{\bf r}_j|)$ \cite{Oseen}.
Substituting this into the above expression for the phase velocity,
we arrive at the coupled phase oscillator equation
\begin{eqnarray}
\dot\phi_i
= \omega_i +
\sum_{j\neq i}
\left(
\frac{{\bf t}_i}{|{\bf r}'_i|}
\cdot \zeta_0{\bf G}_{ij} \cdot |{\bf r}'_j| {\bf t}_j
\right)
\omega_j,
\label{dotphi1}
\end{eqnarray}
where ${\bf G}_{ij} = {\bf G}({\bf r}_i, {\bf r}_j)$.

\subsection{Blake Tensor}

The Blake tensor $\bG(\br_1, \br_2)$ is given by~\cite{Blake}
\eq
G_{\mu\nu}(\br_1, \br_2) &=&
G_{\mu\nu}^S(\br_1 - \br_2)
- G_{\mu\nu}^S(\br_1 - \overline{\br_2})
+ 2 z_2^2 G_{\mu\nu}^D(\br_1 - \overline{\br_2})
- 2 z_2 G_{\mu\nu}^{SD}(\br_1 - \overline{\br_2}),
\label{Blake}
\qe
where
$\br_i = (x_i, y_i, z_i)$ \, $(i=1,2)$,
and $\overline{\br_2} = (x_2, y_2, -z_2)$
is the point of reflection with respect to the substrate,
and
\eq
G_{\mu\nu}^S(\br) &=&
\frac{1}{8\pi\eta}
\left(\frac{\delta_{\mu\nu}}{r} + \frac{r_\mu r_\nu}{r^3}\right),
\\
\nonumber\\
G_{\mu\nu}^D(\br) &=&
\frac{1}{8\pi\eta} \left(1-2\delta_{\nu z}\right)
\frac{\partial}{\partial r_\nu} \left(\frac{r_\mu}{r^3}\right)
\nonumber\\&=&
\frac{1}{8\pi\eta} \left(1-2\delta_{\nu z}\right)
\left(\frac{\delta_{\mu\nu}}{r^3} - \frac{3r_\mu r_\nu}{r^5}\right)
\nonumber\\
&=&
\frac{1}{8\pi\eta}
\left(
\begin{array}{ccc}
\displaystyle
\frac{r^2 - 3x^2}{r^5}
&\displaystyle
\frac{-3xy}{r^5}
&\displaystyle
\frac{-3xz}{r^5}
\\\\\displaystyle
\frac{-3xy}{r^5}
&\displaystyle
\frac{r^2 - 3y^2}{r^5}
&\displaystyle
\frac{-3yz}{r^5}
\\\\\displaystyle
\frac{3xz}{r^5}
&\displaystyle
\frac{3yz}{r^5}
&\displaystyle
\frac{3z^2 - r^2}{r^5}
\end{array}
\right)
,
\label{GD}
\\
\nonumber\\
G_{\mu\nu}^{SD}(\br) &=&
\left(1-2\delta_{\nu z}\right)
\frac{\partial}{\partial r_\nu} G^S_{\mu z}(\br)
\nonumber\\&=&
\frac{1}{8\pi\eta} \left(1-2\delta_{\nu z}\right)
\left(
\frac{\delta_{\mu\nu} r_z + r_\mu \delta_{\nu z}- r_\nu\delta_{\mu z}}{r^3}
- \frac{3 r_{\mu} r_\nu r_z}{r^5}
\right)
\nonumber\\
&=&
\frac{1}{8\pi\eta}
\left(
\begin{array}{ccc}\displaystyle
\frac{z (r^2 - 3x^2)}{r^5}
&\displaystyle
\frac{-3xyz}{r^5}
&\displaystyle
\frac{x (3z^2 - r^2)}{r^5}
\\\\\displaystyle
\frac{-3xyz}{r^5}
&\displaystyle
\frac{z (r^2 - 3y^2)}{r^5}
&\displaystyle
\frac{y (3z^2 - r^2)}{r^5}
\\\\\displaystyle
- \frac{x (r^2 + 3z^2)}{r^5}
&\displaystyle
- \frac{y (r^2 + 3z^2)}{r^5}
&\displaystyle
\frac{z (3z^2 - r^2)}{r^5}
\end{array}
\right)
,
\label{GSD}
\qe
where $\mu,\nu = x,y,z$ with summation over repeated indices assumed,
are the fields of a Stokeslet, source doublet and a Stokeslet doublet,
respectively.
To $O(z_1^2 z_2, z_1 z_2^2)$,
we have
\eq
\bG(\br_1, \br_2)
&=&
\bG(x_1,y_1,z_1, x_2,y_2,z_2)
\nonumber\\
&\simeq&
\frac{3 z_1 z_2}{2\pi\eta \left| \br_1^\perp - \br_2^\perp \right|^5}
\left(
\begin{array}{ccc}
(x_1-x_2)^2
&
(x_1-x_2) (y_1-y_2)
&
-(x_1-x_2) z_2
\\
(x_1-x_2) (y_1-y_2)
&
(y_1-y_2)^2
&
-(y_1-y_2) z_2
\\
(x_1-x_2) z_1
&
(y_1-y_2) z_1
&
0
\end{array}
\right),
\label{Gz3}
\qe
where $\br_i^{\perp} = (x_i, y_i, 0)$
is the horizontal component of the position vector.
Note that $\bG(\br_1,\br_2) \neq \bG(\br_2,\br_1)$
because of
the cubic terms ($G_{3\nu}$, $G_{\mu3}$).
It is convenient to decompose the Blake tensor
into the symmetric and asymmetric part as
\eq
\bG_{12,s}
&=& \frac12 \left[ \bG(\br_1, \br_2) + \bG(\br_2,\br_1) \right]
=
\frac{3 z_1 z_2}{4\pi\eta r_{12\perp}^5}
\left(
\begin{array}{ccc}
2 x_{12}^2
&
2 x_{12} y_{12}
&
x_{12} z_{12}
\\
2 x_{12} y_{12}
&
2 y_{12}^2
&
y_{12} z_{12}
\\
x_{12} z_{12}
&
y_{12} z_{12}
&
0
\end{array}
\right).
\label{G12s}
\\
\bG_{12,a}
&=& \frac12 \left[ \bG(\br_1, \br_2) - \bG(\br_2,\br_1) \right]
=
\frac{3 z_1 z_2}{4\pi\eta r_{12\perp}^5}
\left(
\begin{array}{ccc}
0
&
0
&
-x_{12} w_{12}
\\
0
&
0
&
-y_{12} w_{12}
\\
x_{12} w_{12}
&
y_{12} w_{12}
&
0
\end{array}
\right),
\label{G12a}
\qe
where $\br_{12} = (x_{12}, y_{12}, z_{12}) = \br_1 - \br_2$
and $w_{12} = z_1 + z_2$.

The friction coefficient tensor $\bzeta$ of
the bead at height $z$ from the substrate
is given by~\cite{HappelBrennerBook}
\eq
\zeta_{\mu\nu}(z) = \zeta_0 \left[
\left(1+ \frac{9a}{16z} \right) \delta_{\mu\nu} +
\frac{9a}{16z} \delta_{\mu z}\delta_{\nu z}
\right]
\qquad
\qe
up to $O(a/z)$.
As we stated before,  we assume that the relation $a \ll z$ always holds
and neglect the correction terms.

\subsection{Geometric Factor}

In the following,
we will assume that two rotors have the
same trajectories shape and the force profiles.
We can write each trajectory
as ${\bf r}_i(\phi) = {\bf r}_{i0} + {\bf R}(\phi)$,
where ${\bf r}_{i0}$ is the position of the center, and
${\bf R}(\phi)$ describes the shape of the trajectory.
We also assume that the center positions are
lying along the $x$-axis at height $h$ from the substrate,
and are separated by distance $d$ ($\gg a$) from each other,
and that their coordinates are given by
\eq
{\bf r}_{10} = (0,0,h),
\quad
{\bf r}_{20} = (d,0,h).
\qe
We will also denote the typical size of the trajectory by $b$
and the typical magnitude of the driving force by $F_0$:
\eq
|\bR(\phi)| \sim b, \quad F(\phi) \sim F_0
\qe
Note that
${\bf r}'_i(\phi) = |{\bf R}'(\phi)| {\bf t}(\phi)$
where ${\bf t}(\phi) = {\bf R}'(\phi)/|{\bf R}'(\phi)|$
unit tangential vector of the trajectory.
The intrinsic frequency $\omega(\phi)$ is given by
the force profile $F(\phi)$ as
\begin{eqnarray}
\omega(\phi) = \frac{F(\phi)}{\zeta_0|{\bf R}'(\phi)|}.
\end{eqnarray}
It is useful to rewrite Eq.(\ref{dotphi1})
\eq
\dot{\phi}_i &=& \omega(\phi_i)
\left(
1
+ \sum_{j\neq i} \frac{F(\phi_j)}{F(\phi_i)} H_{ij}(\phi_i, \phi_j)
\right),
\label{dotphi2}
\qe
where
\eq
H_{ij}(\phi_i, \phi_j) &=&
\bt(\phi_i)
\cdot
\zeta_0\bG(\br_i(\phi_i), \br_j(\phi_j))
\cdot
\bt(\phi_j)
\label{Hdef}
\qe
is a dimensionless quantity of $O(a h^2/d^3)$,
and is determined solely by the geometric configuration
of the trajectories (i.e., shape, orientation, distance between
each other, and height from the substrate).
Hereafter we will call $H_{ij}$ the geometric factor.
Note that the symmetry relation
\eq
H_{12}(\phi_1, \phi_2) = H_{21}(\phi_2, \phi_1).
\label{Hsymmetry1}
\qe
holds because $\bG(\br_1, \br_2)$ is identical to
the transposed matrix $\bG^{t}(\br_2, \br_1)$.

\section{Linear Stability Analysis}
\subsection{Generic Conditions for Synchronization}

Let us now examine the stability of the synchronized state
by linearizing the evolution equation of the phase difference
$\delta = \phi_1 - \phi_2$,
which reads
\begin{eqnarray}
\dot{\phi}_1 -  \dot{\phi}_2 &=& \omega(\phi_1) - \omega(\phi_2)
+
\left[
\omega(\phi_1)
\frac{F(\phi_2)}{F(\phi_1)}
-
\omega(\phi_2)
\frac{F(\phi_1)}{F(\phi_2)}
\right]
H_{12}(\phi_1, \phi_2).
\label{dotdelta}
\end{eqnarray}
Here, we used the relation (\ref{Hsymmetry1}).
Setting $\phi_1 = \phi(t) + \delta(t)$, $\phi_2 = \phi(t)$
and linearizing Eq. (\ref{dotdelta}) with respect to $\delta$,
we obtain the linear growth rate
\begin{eqnarray}
\frac{\dot\delta}{\delta}
=
\omega'(\phi) +
\left[
\omega'(\phi) - \frac{2 F'(\phi)}{F(\phi)} \omega(\phi)
\right]
H_{12}(\phi, \phi).
\label{growthrate}
\end{eqnarray}
Integrating (\ref{growthrate}) over the period
$T = \int_0^{2\pi} d\phi/\dot\phi$
in the limit $\delta \to 0$,
we obtain the cycle-averaged growth rate as
\begin{eqnarray}
\Gamma
&=& \frac{1}{T} \int_0^{2\pi} d\phi \,
\frac
{\omega'(\phi) [1+ H_{12}(\phi,\phi)] -2 [\ln F(\phi)]'\omega(\phi) H_{12}(\phi,\phi)}
{ \omega(\phi) [1+ H_{12}(\phi,\phi)]}
\nonumber\\
&\simeq& -\frac{2}{T_0}
\int_0^{2\pi} d\phi \, [\ln F(\phi)]' H_{12}(\phi, \phi),
\label{stabcond}
\end{eqnarray}
where the approximation is taken to the lowest order in the coupling $H_{12}$,
and $T_0$ is the intrinsic period defined by
\eq
T_0 = \int_0^{2\phi} \frac{d\phi}{\omega(\phi)}.
\qe

The synchronized state is stable when $\Gamma<0$.
Equation (\ref{stabcond}) shows
that a necessary condition for synchronization is that both
the force profile $F(\phi)$ and
the geometric factor $H_{12}(\phi,\phi)$ are not constant.
However, the latter is constant only for linear and parallel
trajectories,  as we shall see below.
For other trajectory shapes,
the necessary condition for synchronization
is the non-constantness of the driving force.
Equation (\ref{stabcond}) guarantees that,
if a force profile $F(\phi)$ makes $\Gamma$ positive
for a specific trajectory,
then the force profile proportional to $1/F(\phi)$
makes $\Gamma$ negative for the same trajectory.
In this sense, we can say that roughly  half of
the possible force profiles in the functional space
are capable of inducing in-phase synchronization.
We can also state that,
for any given trajectory shape $\bR(\phi)$ except for
the linear one~\cite{note_linear_trajectory},
there exists a force profile $F(\phi)$
that leads to synchronization.
For example, the force profile
\eq
F(\phi) = F_0 \left[1 + \int_0^\phi d\psi
\left(H_{12}(\psi,\psi) - \overline{H_{12}}\right) \right],
\qe
where $\overline{H_{12}}$ is the period-average of
$H_{12}(\phi,\phi)$, makes $\Gamma$ negative-definite
and hence stabilizes the synchronized state.

\subsection{Far-Field Limit}

Let us now consider the far-field limit
in which the distance between the rotors is
much larger than the typical size of the trajectory.
Also, the height from the substrate is assumed
to be much larger or smaller than the distance:
\eq
\frac{b}{d} \ll 1,
\qquad
\frac{h}{d}
\left\{ \begin{array}{ll} \gg 1 \\ \ll 1 \end{array}
\right.
\qe
In these limits,
the Blake tensor can be approximated by
the sum of isotropic (I) and dyadic (D) parts as
\begin{eqnarray}
\zeta_0{\bf G}_{12}
&\simeq&
G_I(d) {\bf I} + G_D(d) \be_x \be_x
\label{G12FF}
\end{eqnarray}
where we have used $\br_1 - \br_2 = -d \be_x$,
and the dimensionless factors $G_I(d)$ and $G_D(d)$ are given by
\eq
G_I(d) = G_D(d) = \frac{3a}{4d}
\label{Gbulk}
\qe
in the bulk geometry ($h/d \gg 1$) and
\eq
G_I(d) = 0, \quad G_D(d) = \frac{9ah^2}{d^3}
\label{Gsub}
\qe
in the near-substrate geometry ($h/d \ll 1$).
With this approximation, the geometric factor (\ref{Hdef})
reads
\begin{eqnarray}
H_{12}(\phi_1,\phi_2) &=&
G_I(d) +
G_D(d) \, t_x(\phi_1) t_x(\phi_2).
\label{H0th}
\end{eqnarray}
Note that the first term is a constant
and drops off from the integral (\ref{stabcond}).
Therefore, only the non-diagonal part of
the hydrodynamic interaction controls
the stability of the synchronized state in the far-field limit.
Let us examine some specific trajectory/force profiles
in this limit.

\subsubsection{Circular Trajectories}

\begin{figure}[t]
\includegraphics[width=0.74\columnwidth]{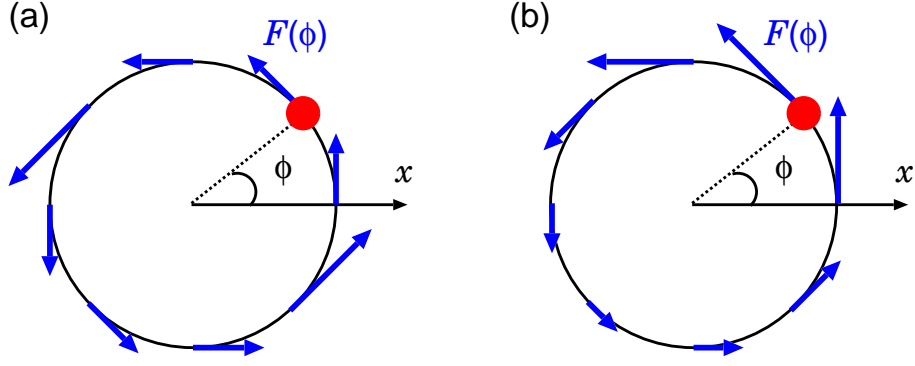}
\caption{
Examples of the force profiles that act to synchronize
two beads on circular trajectories aligned on the $x$-axis.
(a) $F(\phi)=  F_0 [1 - \frac12 \sin (2\phi)]$.
(b) $F(\phi)=  F_0 \left[1 + \frac12
\sin \left(\phi + \frac{\pi}{4}\right)\right]$.
\label{fig:circles}
}
\end{figure}

As the first example, let us consider the circular trajectory
[see Fig. \ref{fig:twobeads}(b)]
\begin{eqnarray}
{\bf R}(\phi) = b (\cos\phi, \sin\phi, 0).
\label{circle}
\end{eqnarray}
For this trajectory,
we have
$|{\bf R}'(\phi)| = b$
and ${\bf t}(\phi) = (-\sin\phi, \cos\phi, 0)$,
which gives
\eq
H_{12}(\phi,\phi) = G_D \sin^2\phi
= -\frac{1}{2} G_D \cos(2\phi) + {\rm const}.
\qe
and
\eq
\Gamma = \frac{G_D}{T_0}\int_0^{2\pi} d\phi \, [\ln F(\phi)]' \cos (2\phi).
\qe
Note that the factor $\cos 2\phi$ represents the second-rank
tensorial nature of the hydrodynamic kernel.
The synchronized state is linearly stable if and
only if the Fourier representation of $\ln F (\phi)$
contains a negative coefficient for $\sin 2\phi$.
Let the Fourier representation of the force profile be
\eq
F(\phi) = F_0 \left[ 1 + \sum_{n=1}^\infty A_n \sin(n\phi + \delta_n) \right]
\label{Ffourier}
\qe
where we assume $0 < A_n < 1$ to avoid singularity of $\ln F$.
Up to $O(A_n^2)$, we obtain the growth rate as
\eq
\Gamma &=& \frac{\pi G_D}{T_0}
\left[
2 A_2 \cos\delta_2 + A_1^2 \sin(2\delta_1) - \sum_{n=1}^\infty
2 A_{n+2} A_n \sin \left(\delta_{n+2} - \delta_n \right)
\right].
\label{Gammacircle}
\qe
The only harmonic mode that contributes to $\Gamma$ at $O(A_n)$
is $n=2$, for which $\Gamma$ is most negative at $\delta_2 = \pi$.
In this sense, the force profile that is most efficient
in inducing synchronization is given by
\eq
F(\phi)=  F_0 [1 - A_2 \sin (2\phi)], \quad 0<A_2 < 1.
\label{Fsin2phi}
\qe
Another harmonic mode that contributes to $\Gamma$ by itself is $n=1$,
for which $\Gamma$ is most negative at $\delta_1 = {\pi}/{4}, {5\pi}/{4}$.
Thus we obtain the candidate force profile
\eq
F(\phi)=  F_0 \left[1 + A_1 \sin \left(\phi + \frac{\pi}{4}\right)\right],
\qquad
-1 < A_1 < 1.
\label{Fsinphi}
\qe
These two force profiles are illustrated in Fig. \ref{fig:circles}.
Note that higher harmonic modes ($n\ge 3$)
can stabilize synchronization
only when they are mixed with the other modes.

Next, we consider rotated circular trajectories.
The  trajectory (\ref{circle})  rotated
around the $y$- axis by angle $\alpha$,
$
\bR(\phi) = b(\cos\alpha \cos \phi, \sin\phi, \sin\alpha \cos\phi)
$,
gives the additional factor $\cos^2 \alpha$
to the linear growth rate via Eq.(\ref{H0th}).
Note that if the trajectory planes are perpendicular
to the $x$-axis ($\alpha=\pi/2$),
we have $\Gamma=0$ and the synchronized state is only marginally stable.
On the other hand,
rotation around the $x$-axis does not change the linear growth rate,
because the hydrodynamic kernel (\ref{G12FF}) is invariant for the rotation.
However, near-field corrections will introduce an important dependence,
as we shall see in the next section.
We do not consider rotation around the $z$-axis, which is
equivalent to shift of the phase by a constant.

\subsubsection{Linear Trajectories}

The linear trajectory
\eq
{\bf R}(\phi) = R(\phi) {\bf e}_x,
\label{linear}
\qe
gives ${\bf t}(\phi) = {\rm sgn} [R'(\phi)] {\bf e}_x$,
which makes the geometric factor (\ref{H0th}) constant.
Thus, at the level of linear stability analysis,
the synchronized state is neither stabilized nor destabilized
for any force profile.
However, nonlinear stability analysis shows
that the stability is weakly affected by force modulation,
as we shall see in the next section.

\subsubsection{Elliptic Trajectories}

For the elliptic trajectory
\eq
\bR(\phi) = (b_x \cos\phi, b_y \sin\phi), \quad b_x,b_y>0,
\label{ellipse}
\qe
the $x$-component of the tangential vector
$t_x(\phi) = b_x\cos\phi/\sqrt{b_x^2 \cos^2 \phi
+ b_y^2 \sin^2\phi}$
contains all the harmonic modes with odd $n$ if $b_x \neq b_y$.
As a result, force modulations containing any harmonic mode
with even $n$ can induce synchronization at $O(A_n)$,
if the Fourier coefficients are suitably chosen.
For example, when $b_x > b_y$,
the force profile
\eq
F(\phi) = F_0 \left[ 1 + A_4 \sin (4\phi) \right], \quad 0 < A_4 < 1
\label{Fsin4phi}
\qe
gives the negative growth rate up to $O(A_4)$,
\eq
\Gamma = -\frac{\pi G_D A_4}{2T_0} \, \chi (1+\chi)(2+\chi), \qquad
\chi = \frac{b_x^2-b_y^2}{b_x^2+b_y^2}.
\qe
Force modulations with odd harmonics can also induce synchronization,
because they give rise to even harmonic modes in $\ln F(\phi)$,
but only at $O(A_n^2)$.

\section{Flow Rate and Energy Dissipation}

Now let us see how synchronization affects flow properties
in the substrated geometry.
We define the volume flow rate $Q$ as the flux through
a half-plane in the ``down stream'' ($x\to-\infty$):
\eq
Q(t) &=&
- \lim_{x\to-\infty}
\int_{-\infty}^{\infty} dy \int_0^\infty dz \,
v_x(\br, t)
\nonumber\\
&=&
\lim_{x\to-\infty}
\int_{-\infty}^{\infty} dy \int_0^\infty dz \,
\sum_i G_{x\nu}(\br, \br_i) \cdot \bg_i.
\label{flowrate1}
\qe
In the second line we used the expression for the flow field (\ref{vflow}).
Note that, due to volume conservation,
the integral in (\ref{flowrate1}) does not depend on the $x$-position
of the half-plane. However, it is easier to calculate
it in the limit $x\to-\infty$,  where we can use the
$O(r^{-2})$ approximation for the Blake tensor,
\eq
G_{\mu\nu}(\br, \br_i) =
\frac{3z z_i}{2\pi\eta r^5}
\left(
\begin{array}{ccc}
\displaystyle
x^2
&\displaystyle
xy
&\displaystyle
xz
\\\displaystyle
xy
&\displaystyle
y^2
&\displaystyle
yz
\\\displaystyle
0
&\displaystyle
0
&\displaystyle
0
\end{array}
\right),
\qe
which gives 
\eq
Q(t) &=&
\frac{1}{\pi\eta}
\sum_i [h + R_z(\phi_i)] \, g_{ix}.
\label{flowrate2}
\qe
Because we are interested in the change in the flow rate
due to hydrodynamic interaction between the rotors,
we retain the first order term with respect to $\bG$
in calculating the drag force,
which reads,
\eq
\bg_i &=& \zeta_0 [\bv(\br_i) - \dot{\br}_i]
\nonumber\\
&=&
\zeta_0
\left[
\sum_{j\neq i} \zeta_0 \bG(\br_i, \br_j) \cdot \bR'(\phi_j) \dot{\phi}_j 
-
\bR'(\phi_i) \dot{\phi}_i
\right].
\label{gexplicit}
\qe
In the in-phase synchronized state $\phi_i = \phi_j = \phi$,
the two rotors have the same period $T$, and
the cycle-averaged flow rate is calculated using 
(\ref{flowrate2}) and (\ref{gexplicit})
as
\eq
\overline{Q} &=& \frac{1}{T} \int_0^T dt \, Q(t)
\nonumber\\
&=&
-\frac{6a}{T}
\sum_i
\int_0^{2\pi} d\phi
\left[ h+R_z(\phi) \right]
\left[
R_{x}'(\phi)
-
\sum_{j\neq i}
\zeta_0 G_{x\nu}(\br_i, \br_j) R'_{\nu}(\phi)
\right].
\label{meanflowrate1}
\qe
We shall use the far-field approximation (\ref{G12FF}) and (\ref{Gsub}), 
to obtain
\eq
\overline{Q} = \frac{12a (1- G_D)}{T}
\int_0^{2\pi} d\phi\, R_z(\phi) R_{x}'(\phi).
\label{meanflowrate2}
\qe
Note that the flow rate is zero for planar geometry ($R_z(\phi)=0$).
The hydrodynamic interaction modifies the flow rate 
not only through the prefactor $1 - G_D$
but also through the period $T$, which is given by
\eq
T = \int_0^{2\pi} \frac{d\phi}{\dot\phi} \,
\simeq
\int_0^{2\pi} \frac{d\phi}{\omega(\phi)}
\left[ 1- H_{12}(\phi, \phi) \right].
\label{period}
\qe

We can also calculate the power needed to drive the beads,
which is given by
\eq
P(t) = \sum_i \dot{\br}_i \cdot (-\bg_i).
\label{power}
\qe
Its cycle-average in the synchronized state
is calculated to the first order of $\bG$
as
\eq
\overline{P} &=&
\frac{\zeta_0}{T}
\sum_i
\int_0^{2\pi} d\phi
\left[
\dot{\phi} |\bR'(\phi)|^2
- \omega(\phi) \bR'(\phi) \cdot
\sum_{j\neq i} \zeta_0 \bG(\br_i, \br_j) \cdot \bR'(\phi)
\right]
\nonumber\\
&=&
\frac{2\zeta_0}{T}
\int_0^{2\pi} d\phi \, \omega(\phi)
\left\{
\left[ 1+H_{12}(\phi,\phi) \right] |\bR'(\phi)|^2
- G_D R_x'(\phi)^2
\right\}.
\label{meanpower2}
\qe

For example,
let us compute the flow rate and power
for the vertical circular trajectory
\eq
\bR(\phi) = b(\cos\phi, 0, \sin\phi).
\qe
For this trajectory,
the integrals in (\ref{meanflowrate2}) and (\ref{meanpower2})
give $\pi b^2$ and $2\pi F_0 b/\zeta_0$, respectively,
where $F_0$ is the cycle-average of the driving force.
It yields the mean flow rate
\eq
\overline{Q} = \frac{12\pi a b^2 (1- G_D)}{T}
\label{meanflowrate3}
\qe
and the mean power
\eq
\overline{P} = \frac{4\pi F_0 b}{T}.
\label{meanpower3}
\qe
The period $T$ depends on the force profile,
and is given by
\eq
T &=& \frac{2\pi \zeta_0 b}{F_0} (\tau_0 - \tau_1 G_D),
\\
\tau_0 &=&
\frac{1}{2\pi} \int_0^{2\pi} d\phi \, \frac{F_0}{F(\phi)},
\\
\tau_1 &=&
\frac{1}{2\pi} \int_0^{2\pi} d\phi \, \frac{F_0 \sin^2\phi}{F(\phi)}.
\label{tau}
\qe
The dimensionless coefficients $\tau_0$ and $\tau_1$
are positive for any force profile (with $F(\phi)>0$),
which means that the period decreases
by the hydrodynamic interaction.
Furthermore, we have $\tau_1/\tau_0 < 1$
for any force profile,
which means that the mean flow rate also decreases
by the hydrodynamic interaction.

\begin{figure}[t]
\includegraphics[width=0.49\columnwidth]{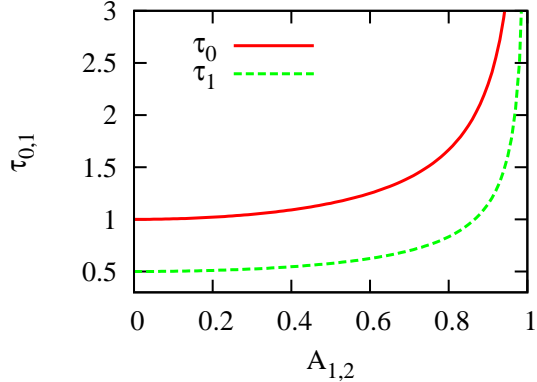}
\caption{
The coefficients $\tau_0$ and $\tau_1$ in Eq.(\ref{tau})
for the force profiles (\ref{Fsin2phi}) and (\ref{Fsinphi}),
as functions of $A_1$ and $A_2$ (resp.).
For the two force profiles, the curves are identical
and $\tau_1/\tau_0$ is equal to $1/2$.
\label{fig:tau}
}
\end{figure}

In Fig. \ref{fig:tau}, we plot $\tau_0$ and $\tau_1$
for the force profiles (\ref{Fsin2phi}) and (\ref{Fsinphi})
as functions of the amplitude $A_2$ and $A_1$,
respectively.
The dependencies on $A_1$ and $A_2$ turn out to be identical
for each of $\tau_0$ and $\tau_1$,
with the ratio $\tau_1/\tau_0$ equal to $0.5$.
Both of the coefficients, and hence the period,
diverge as $A_{1,2}$ are taken to unity.
When $A_{1,2}=1$, there are stall points
(where $F(\phi)=0$) on the trajectory,
and it takes infinite time for the bead to pass the points.

\section{Nonlinear Analysis}

In this section, we analyze the fully nonlinear evolution equation
for the phase difference (\ref{dotdelta}),
which allows us to explore various dynamical states and their stability.
By using the full Blake tensor,
we will also discuss the near-field effects due to finite size
and height of the trajectories.

\subsection{Effective Potential}
The difference in the phase velocities (\ref{dotdelta})
consists of the intrinsic phase velocities (the first and second terms in the RHS)
and the interaction term (the third term in the RHS).
In order to focus on the latter, we exploit
the gauge invariance of Eq.(\ref{dotdelta}), i.e.
the invariance under the transformation $\phi \to \Phi(\phi)$ where
$\Phi$ is a new phase variable (or a ``gauge'')
satisfying $\Phi(\phi+2\pi) = \Phi(\phi) + 2\pi$.
We choose the specific gauge $\Phi$ that gives a
constant intrinsic phase velocity, which we will call the canonical gauge.
It satisfies $\dot\Phi = 2\pi/T_0 = \Omega$ in the absence of
hydrodynamic interaction, and is obtained from the original gauge $\phi$
via the relation
\eq
\frac{d\Phi}{d\phi} = \frac{\dot\Phi}{\dot\phi}
= \Omega \frac{\zeta_0 |\bR'(\phi)|}{F(\phi)}.
\label{phitoPhi}
\qe
In the canonical gauge, the intrinsic terms in Eq.(\ref{dotdelta}) cancel out,
and the phase difference $\Delta = \Phi_1 - \Phi_2$ obeys
\eq
\dot\Delta = \Omega
\left[
\frac{\tilde{F}(\Phi_2)}{\tilde{F}(\Phi_1)} -
\frac{\tilde{F}(\Phi_1)}{\tilde{F}(\Phi_2)}
\right]
\tilde{H}_{12}(\Phi_1, \Phi_2),
\label{dotDelta1}
\qe
where the force profile $\tilde{F}(\Phi)$ and the geometric factor
$\tilde{H}_{12}(\Phi_1, \Phi_2)$ are related to those in the original gauge
via $\tilde{F}(\Phi) = F(\phi)$ and
$\tilde{H}_{12}(\Phi_1, \Phi_2) = H_{12}(\phi_1, \phi_2)$.
Note also that $\tilde{\bR}(\Phi) = \bR(\phi)$
and $\tilde{\bR}'(\Phi) = \frac{d\phi}{d\Phi} \bR'(\phi)
= \frac{1}{\Omega\zeta_0} F(\phi) \bt(\phi)$.
We rewrite (\ref{dotDelta1}) in terms of $\Delta$ and the phase sum
$\Sigma = \Phi_1 + \Phi_2$, as
\eq
\dot\Delta =
\Omega
\left[
\frac
{\tilde{F}\left(\frac{\Sigma-\Delta}{2}\right)}
{\tilde{F}\left(\frac{\Sigma+\Delta}{2}\right)}
-
\frac
{\tilde{F}\left(\frac{\Sigma+\Delta}{2}\right)}
{\tilde{F}\left(\frac{\Sigma-\Delta}{2}\right)}
\right]
\tilde{H}_{12}
{\textstyle 
\left(\frac{\Sigma+\Delta}{2},\frac{\Sigma-\Delta}{2}\right)
}
=
W(\Sigma, \Delta).
\label{Udef}
\qe
Note that $\dot{\Delta}/\Omega = O(G_D) \ll 1$,
while $\dot{\Sigma}/\Omega$ is an $O(1)$ quantity.
Therefore,
we can approximate $\Delta$ to be constant over one period
where $\Sigma$ increases by $4\pi$.
With the approximation,
we take the average of (\ref{Udef})
over one period $0<t<T$, to obtain
\eq
\dot{\Delta}
= \frac{1}{4\pi} \int_0^{4\pi} d\Sigma \, \, W(\Sigma, \Delta)
= \overline{W}(\Delta),
\label{Wdef}
\qe
which defines the effective force $\overline{W}(\Delta)$.
We introduce the effective potential $V(\Delta)$ by
\eq
V(\Delta) = - \int_0^\Delta d\Delta' \overline{W}(\Delta')
\label{Vdef}
\qe
with which the dynamics reduces to minimizing the potential:
\eq
\dot\Delta = -\frac{dV}{d\Delta}.
\label{dotDelta3}
\qe
Thus we have eliminated the fast variable $\Sigma$ and
describe the slow dynamics only by $\Delta$.
This approximation is shown to be to correct to the lowest order
in the interaction~\cite{Kuramoto}.

Note that the factor
$
{\tilde{F}\left(\frac{\Sigma-\Delta}{2}\right)}/
{\tilde{F}\left(\frac{\Sigma+\Delta}{2}\right)}
-
{\tilde{F}\left(\frac{\Sigma+\Delta}{2}\right)}/
{\tilde{F}\left(\frac{\Sigma-\Delta}{2}\right)}
$
in Eq. (\ref{Udef}) is an odd function of $\Delta$ by construction.
The other factor
$
\tilde{H}_{12}
{
\left(\frac{\Sigma+\Delta}{2},\frac{\Sigma-\Delta}{2}
\right)
}
$
is an even function of $\Delta$ if the following identity holds:
\eq
\tilde{H}_{12}(\Phi_1,\Phi_2) = \tilde{H}_{12}(\Phi_2,\Phi_1).
\label{Hsymmetry2}
\qe
This is the case for the far-field limit ($b/d \to 0$),
as we see from Eqs.(\ref{Hdef}) and (\ref{G12FF}).
In that case, $W(\Sigma, \Delta)$ and $\overline{W}(\Delta)$ are odd
functions of $\Delta$ and hence the effective potential is
an even function: $V(\Delta) = V(-\Delta)$.

\subsection{Far-Field Limit}
First, let us derive the effective potential in the far-field limit
($b/d \to 0$),
for trajectories in the bulk ($h/d \to \infty$) and/or
near substrate ($h/d \ll 1$).

\subsubsection{Circular Trajectories}

\begin{figure}
\includegraphics[width=0.99\textwidth]{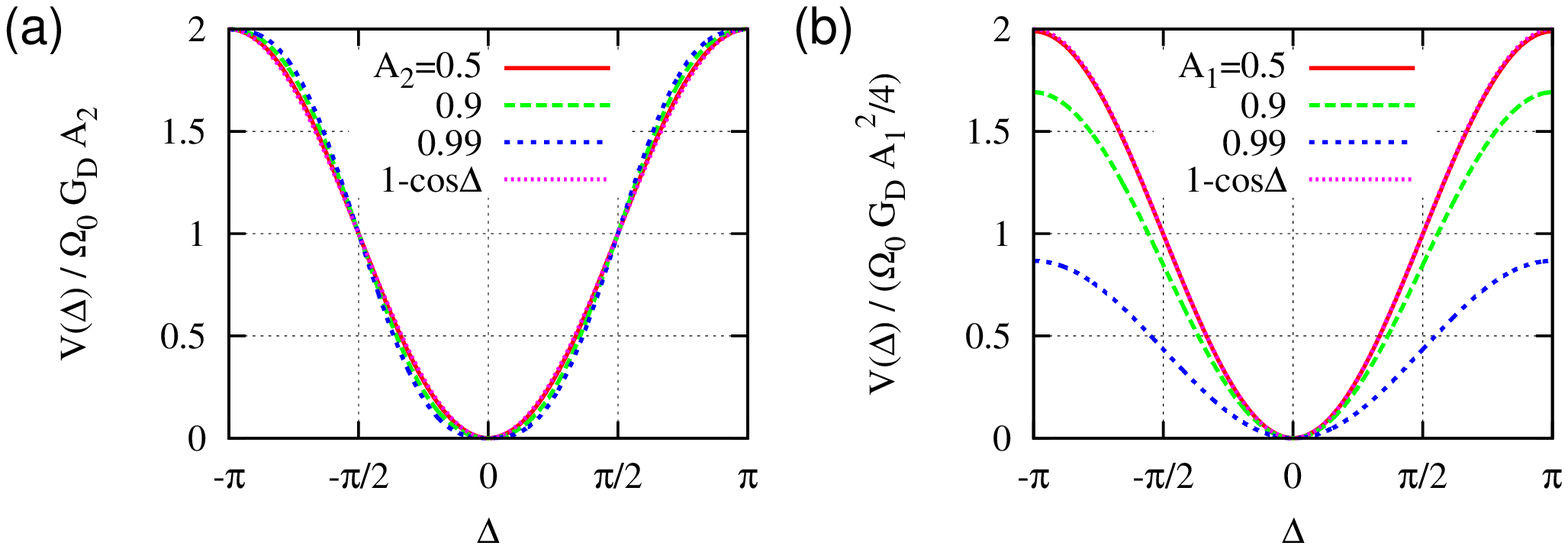}
\caption{
The effective potential $V(\Delta)$ in the far-field limit $b/d \to 0$
for the circular trajectory (\ref{circle}), either in the bulk
($h/d \gg 1$) or near the substrate ($h/d \ll 1$),
for (a) the force profile (\ref{Fsin2phi}),
and (b) the force profile (\ref{Fsinphi}).
}
\label{fig:A2dep-0}
\end{figure}

First let us consider circular trajectory (\ref{circle})
with the force profile (\ref{Fsin2phi}).
The phase in the canonical gauge is obtained via (\ref{phitoPhi}), as
\eq
\Phi(\phi) &=& 2\pi \cdot \frac{K(\phi)}{K(2\pi)},
\quad
K(\phi) = \int_0^\phi \frac{d\phi'}{1 - A_2 \sin 2\phi'}.
\label{Phi_sin2phi}
\qe
Accordingly, the intrinsic phase velocity in the canonical gauge is given by
\eq
\Omega &=& \frac{\Omega_0}{K(2\pi)},
\quad
\Omega_0 = \frac{2\pi F_0}{\zeta_0b}.
\label{Omega_sin2phi}
\qe
For $A_2\ll1$, we can approximate
$\phi(\Phi)$ and $\tilde{F}(\Phi)$ as
$\phi = \Phi + \frac{A_2}{2} \cos 2\Phi$
and
$\tilde{F}(\Phi) = F(\Phi)$
to $O(A_2)$,
which gives
\eq
V(\Delta) \simeq V_0(\Delta) = \Omega_0 G_D A_2 (1-\cos \Delta).
\label{Vsin2phi}
\qe
For not small values of $A_2$, we compute the integrals
in (\ref{Wdef}) and (\ref{Phi_sin2phi}) numerically.
We plot $V(\Delta)$ in Fig.\ref{fig:A2dep-0},
which shows that the approximation (\ref{Vsin2phi}) is very good.
Even for $A_2=0.99$,
the deviation $[V(\Delta)-V_0(\Delta)]/V_0(\Delta)$
falls within $11$ \% for any value of $\Delta$.
Also note that $V(\Delta)$ is an odd function of $A_2$.
For $A_2 > 0$, it is minimized at the in-phase synchronized state ($\Delta=0$),
while for $A_2 < 0$, it is minimized at the anti-phase synchronized state
($\Delta= \pi$).

For the force profile (\ref{Fsinphi}),
the effective potential can be calculated in a similar way as above,
and is plotted in Fig. \ref{fig:A2dep-0}.
A perturbative calculation to $O(A_1^2)$ gives
\eq
V(\Delta) \simeq V_0(\Delta)
= \frac{\Omega_0 G_D A_1^2}{4} (1-\cos \Delta).
\label{Vsinphi}
\qe
Again, the approximation (\ref{Vsinphi}) is good for
moderate values of $A_1$.
The deviation $[V(\Delta) - V_0(\Delta)]/V_0(\Delta)$ falls
within $0.02$ for $A_1=0.5$.
Although the deviation increases to $0.57$ at $A_1=0.99$,
the shapes of $V(\Delta)$ and $V_0(\Delta)$ are quite similar.

\subsubsection{Linear Trajectories}

\begin{figure}
\includegraphics[width=0.48\textwidth]{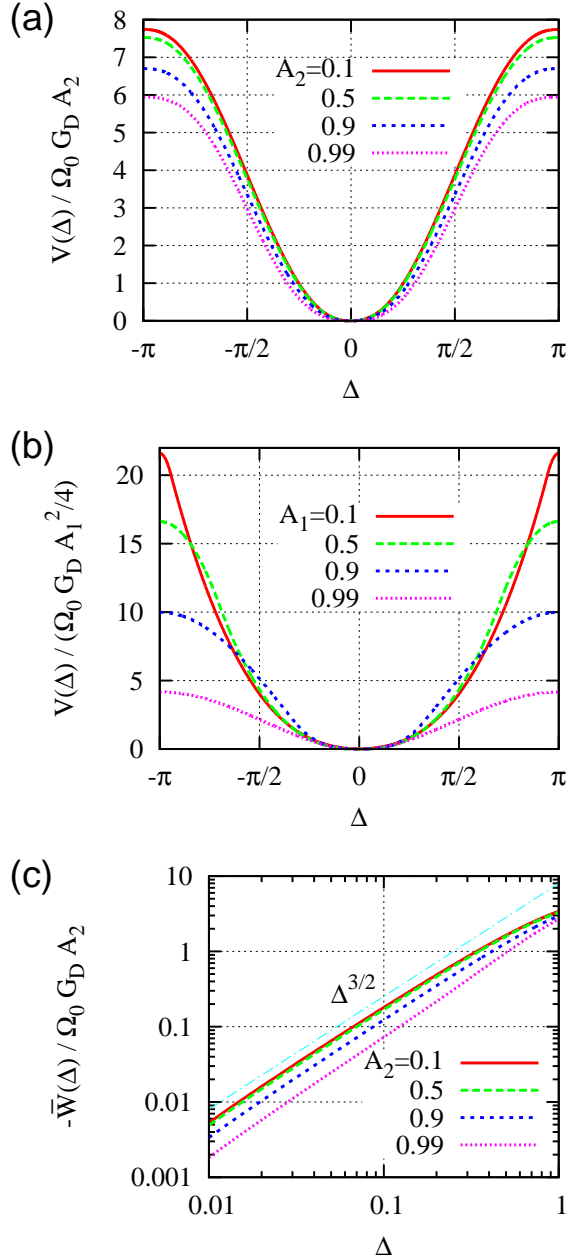}
\caption{
The effective potential $V(\Delta)$ in the far-field limit $b/d \to 0$
for the linear trajectory (\ref{circle})
near the substrate ($h/d \ll 1$),
for
(a) the force profile (\ref{Fsin2phi})
and
(b) the force profile (\ref{Fsinphi}).
(c) Non-analytic behavior
$V'(\Delta) = -\overline{W}(\Delta) \propto \Delta^{3/2}$
for the case (a).
}
\label{fig:linear}
\end{figure}

Next let us consider the linear trajectory (\ref{linear}) with
$R(\phi) = b \cos \phi$ using the near-substrate approximation.
In Fig. \ref{fig:linear},
we plot the effective potential
for (a) the force profile (\ref{Fsin2phi}) and
(b) the force profile (\ref{Fsinphi}).
In both cases the potential is minimized at $\Delta=0$.
This result is not expected from the linear stability analysis,
which showed that the in-phase synchronized state
is only marginally stable for any force profile.
It suggests that the potential scales with
$V(\Delta) \propto |\Delta|^{\kappa}$ with
the exponent $\kappa > 2$ near $\Delta=0$.
In Fig.\ref{fig:linear}(c), we plot
$V'(\Delta)=-\overline{W}(\Delta)$ for the case (a).
It indicates the non-analytical behavior $\kappa=5/2$,
which we will prove in the following paragraph.

The gauge condition (\ref{phitoPhi}) gives
\eq
\Phi(\phi) &=& 2\pi \cdot \frac{K(\phi)}{K(2\pi)},
\qquad
K(\phi) = \int_0^\phi d\phi' \frac{|\sin\phi'|}{1 - A \sin 2\phi'}
\label{Phiphi}
\qe
We see that $\Phi(\pi + \phi) = \pi + \Phi(\phi)$, and especially
$\Phi(\pi) = \pi$.
For $|\phi| \ll 1$, we have $K(\phi) \approx {\rm sgn(\phi)} \cdot \phi^2/2$
and hence $\Phi(\phi) \approx [\pi/K(2\pi)] \,{\rm sgn(\phi)} \cdot \phi^2$.

We also expand the factor in (\ref{Udef}) in powers of $\Delta$ as
\eq
\frac
{\tilde{F}\left(\frac{\Sigma-\Delta}{2}\right)}
{\tilde{F}\left(\frac{\Sigma+\Delta}{2}\right)}
-
\frac
{\tilde{F}\left(\frac{\Sigma+\Delta}{2}\right)}
{\tilde{F}\left(\frac{\Sigma-\Delta}{2}\right)}
=
- \Delta \frac{d}{d\Sigma} 
\left[ 
\ln \tilde{F} \left(\frac{\Sigma}{2}\right) 
\right]
+ O(\Delta^3)
\label{FoverFtaylor}
\qe
while the other factor behaves like a step function:
\eq
\tilde{H}_{12}\left(\frac{\Sigma+\Delta}{2},\frac{\Sigma-\Delta}{2}\right)
&=&
G_D \, {\rm sgn} \left[
\sin\left(\frac{\Sigma+\Delta}{2}\right)
\sin\left(\frac{\Sigma-\Delta}{2}\right)
\right].
\qe
For small and positive value of $\Delta$, the latter equals
$-G_D$ when $2n \pi - \Delta < \Sigma < 2n\pi + \Delta$
($n$: integer) and equals $G_D$ otherwise.
These give the effective force to $O(\Delta^2)$ as
\eq
\overline{W}(\Delta) \approx
\frac{\Omega G_D \Delta}{2\pi}
\ln
\frac
{\tilde{F}\left( \frac{\Delta}{2}\right)}
{\tilde{F}\left(-\frac{\Delta}{2}\right)}
\approx
-\frac{\Omega G_D A}{2\pi}  \sqrt{\frac{K(2\pi)}{\pi} |\Delta|} \cdot \Delta,
\qe
where we used $\ln \tilde{F}(\Phi) = \ln F(\phi)
\approx \ln F_0 - 2 A \phi
\approx \ln F_0 - {\rm sgn}(\Phi) \cdot A \sqrt{[K(2\pi)/\pi] |\Phi|}$,
which is an approximation for $|\Phi| \ll 1$.
Thus we obtained the non-analytic behavior
$\overline{W}(\Delta) \propto -{\rm sgn}(\Delta) \cdot |\Delta|^{3/2}$, or
$V(\Delta) \propto  {\rm sgn}(\Delta) \cdot |\Delta|^{5/2}$.

\subsubsection{Elliptic Trajectories}

\begin{figure}
\includegraphics[width=0.99\textwidth]{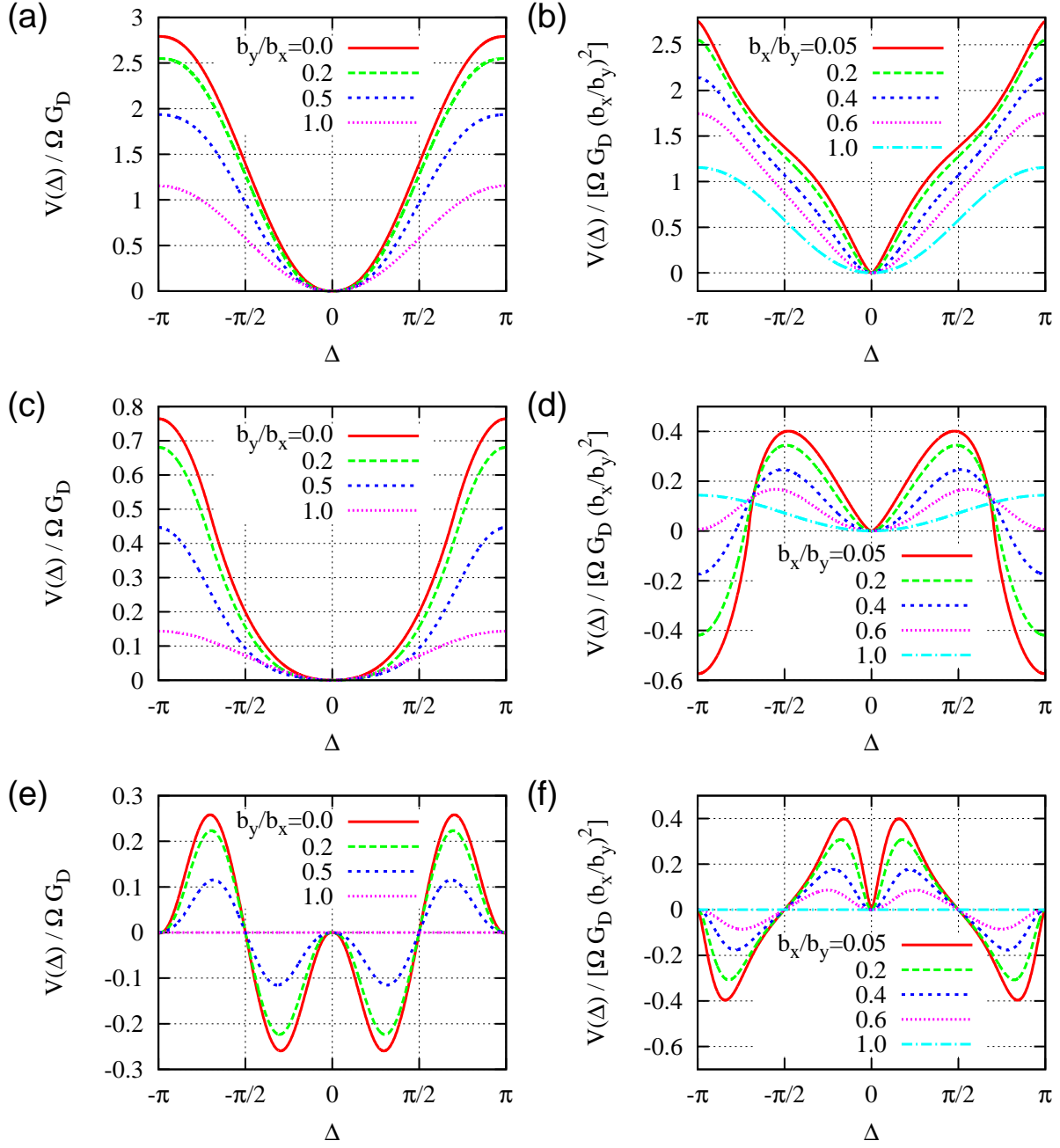}
\caption{
The effective potential $V(\Delta)$ in the far-field limit $b/d \to 0$
for the elliptic trajectory (\ref{ellipse})
near the substrate ($h/d \ll 1$),
for
(a,b) the force profile (\ref{Fsin2phi}) with $A_2=0.5$,
(c,d) the force profile (\ref{Fsinphi}) with $A_1=0.5$,
and
(e,f) the force profile (\ref{Fsin4phi}) with $A_4=0.5$.
The long-axis of the ellipse is along the $x$-direction in (a,c,e),
and along the $y$-direction in (b,d,f).
\label{fig:ellipse}
}
\end{figure}

Next we consider the elliptic trajectory (\ref{ellipse})
in the near-substrate approximation.
For elliptic trajectories, the tangential vector
$\bt(\phi) = \bR'(\phi)/|\bR'(\phi)|$
and hence the geometric factor (\ref{H0th})
contain various harmonic modes, which produce
richer behaviors than the circular trajectories.
In Fig. \ref{fig:ellipse}, we show the effective potential
for
(a,b) the force profile (\ref{Fsin2phi}) with $A_2=0.5$,
(c,d) the force profile (\ref{Fsinphi}) with $A_1=0.5$,
and
(e,f) the force profile (\ref{Fsin4phi}) with $A_4=0.5$.
In (a,c,e), we show the potential curves for $b_x \ge b_y$,
while in (b,d,f), the potential curves for $b_x \le b_y$
are scaled by $(b_x/b_y)^2$.
(Note that the potential converges to zero in the limit
$b_x/b_y \to 0$.)

In (a), the potential has a single minimum at $\Delta=0$.
As $b_x/b_y \to 0$, the scaled potential $V(\Delta)/(b_x/b_y)^2$
converges to a V-shape curve.

In (b), a local minimum at $\Delta=\pm\pi$ appears for $b_x/b_y<1$
in addition to the minimum at $\Delta=0$.
For $b_x/b_y \simle 0.6$, the anti-phase synchronized state
becomes stable, while the in-phase state becomes metastable.
For $\delta_2=3\pi/4$, the sign of the potential has an opposite sign,
and we obtain bistable minima at $\Delta=\pm \Delta_0$,
with $\Delta_0 \simeq \pi/2$ for $b_x/b_y=0.4$ and
$\Delta_0 \to \pi$ as $b_x/b_y$ is increased to unity.

In (c), we have bistable minima at $\Delta \simeq \pm \Delta_0$
with $\pi/2<\Delta_0<\pi$ for $b_x/b_y<1$,
and  $0<\Delta_0<\pi/2$ for $b_x/b_y>1$.
A metastable minimum is located at $\Delta = 0$ and $\Delta=\pm\pi$,
for $b_x/b_y<1$ and $b_x/b_y>1$, respectively.
For $b_x/b_y=1$, no phase locking occurs because
the potential is constant.
In (b) and (c) also, the scaled potential $V(\Delta)/(b_x/b_y)^2$
converges to a master curve with sharp peaks and valleys
in the limit $b_x/b_y \to 0$ (not shown).

\begin{figure}[t]
\includegraphics[width=0.99\textwidth]{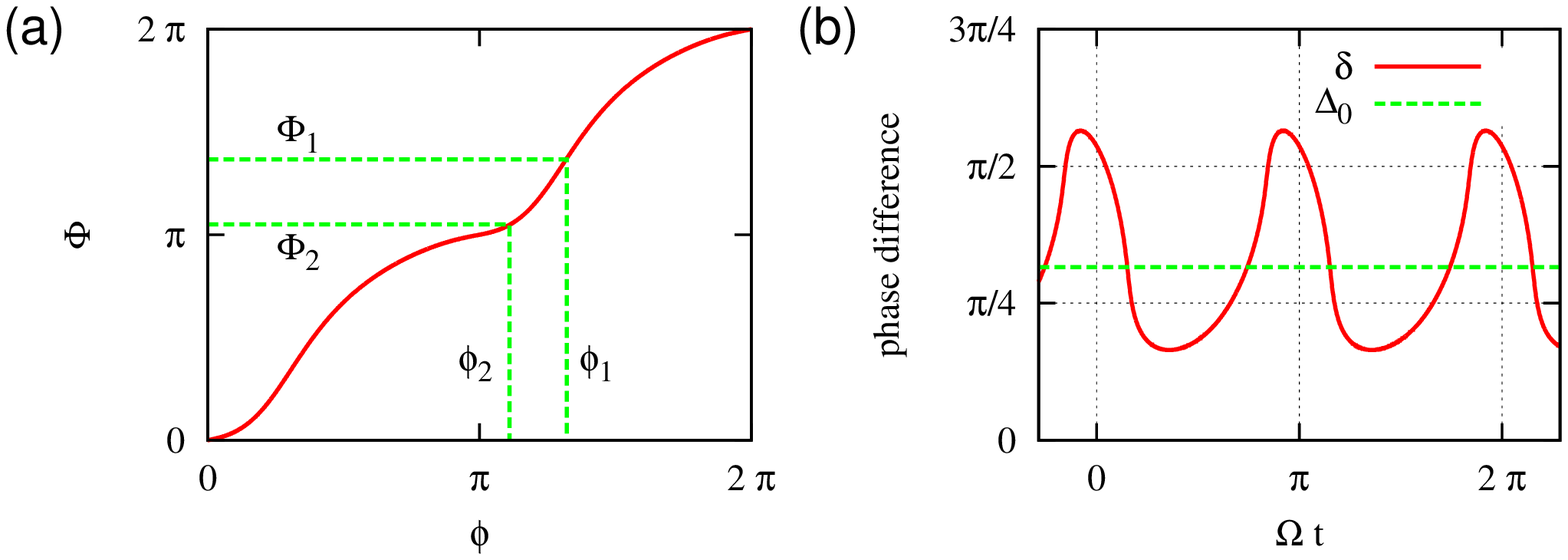}
\caption{
(a) The phase function $\Phi(\phi)$
for the elliptic trajectory with $b_y/b_x = 0.5$
and the force profile (\ref{Fsin4phi}) with $A_4=0.5$.
(b) Temporal oscillation of $\delta$
in the phase-locked state $\Delta=\Delta_0=0.992$.
\label{fig:delta}
}
\end{figure}

Now let us consider the meaning of the minimum at non-zero $\Delta$.
In the phase-locked state, the phase difference $\Delta$
in the canonical gauge is constant, but it generally means
an oscillation of the phase difference $\delta$
in the original gauge,
because it is a function of both $\Delta$ and $\Sigma \simeq 2\Omega t$.
Let us take for example, the elliptic trajectory with
$b_y/b_x = 0.5$ and the force profile (\ref{Fsin4phi}) with $A_4=0.5$.

In Fig. \ref{fig:delta}(a) we show the phase function $\Phi(\phi)$,
which has the period $\pi$.
The figure also illustrates the relation between $\delta=\phi_1-\phi_2$
and $\Delta=\Phi_1-\Phi_2$.
The effective potential, shown in Fig.\ref{fig:ellipse}(e),
has double minima at $\Delta=\pm \Delta_0$ with $\Delta_0 = 0.992$.

Also, it has a metastable minimum at $\Delta=\pi$.
In Fig. \ref{fig:delta}(b), we show $\delta$
in the phase-locked state $\Delta=\Delta_0$
as a function of time
(using the relation $\Omega t = \Sigma/2$,
where the origin of time is chosen arbitrarily).
It oscillates with the period of $\Phi(\phi)$.
The amplitude of oscillation is as large as $1.25$,
and is larger for stronger modulation of the
force profile (that is, for larger amplitude $A_4$).
On the other hand, $\delta$ remains constant in the
anti-phase synchronized state $\Delta=\pi$,
because it matches the period of the phase function.

\subsection{Near-Field Corrections}

Next we consider the near-field effects
arising from finite size of the trajectory,
by using the full Blake tensor $\bG_{12}$ given by (\ref{Blake}).
The finite size of the trajectory introduces dependences of $\bG_{12}
=\bG_{12}(\br_{10} + \tilde{\bR}(\Phi_1), \br_{20} + \tilde{\bR}(\Phi_2))$
on $\Phi_1$ and $\Phi_2$. In general, $ \bG_{12}$ is not symmetric
with respect to the exchange of $\Phi_1$ and $\Phi_2$, which leads
to asymmetry of the effective potential $V(\Delta)$.
To see this explicitly, we expand the Blake tensor
to the first order with respect to
\eq
\brho_i = \frac{\tilde\bR(\Phi_i)}{d}, \qquad i=1,2,
\qe
which is assumed to be small.
We also assume that the height of the trajectory
is of the same order as its size, and introduce
the dimensionless height,
\eq
\hat{h} = \frac{h}{d}.
\qe
Then we can use the $O(\hat{h}^3)$ approximation (\ref{Gz3})
as a starting point for the expansion.
Substituting
$x_{12} = d(-1 + \rho_{1x} - \rho_{2x}), \,
y_{12} = d (\rho_{1y} - \rho_{2y}),\,
z_{i} = d (\hat{h} + \rho_{iz}), \,
z_{12} = d(\rho_{1z} - \rho_{2z}),$
and
$w_{12} = d(2 \hat{h} + \rho_{1z} + \rho_{2z})$
into (\ref{G12s},\ref{G12a})
and retaining $O(\brho, \hat{h})$ terms, we have
\eq
\zeta_0 \bG_{12,s}^{(2)}
&\simeq&
C_D 
\left(1+\frac{\rho_{1z}}{\hat{h}} \right) 
\left(1+\frac{\rho_{2z}}{\hat{h}} \right)
\left(
\begin{array}{ccc}
1+3(\rho_{1x} - \rho_{2x})
&
-(\rho_{1y} - \rho_{2y})
&
0
\\
-(\rho_{1y} - \rho_{2y})
&
0
&
0
\\
0
&
0
&
0
\end{array}
\right)
\nonumber\\ &&
+
\frac{C_D \hat{h}}{2}
\left(1+\frac{\rho_{1z}}{\hat{h}}\right)
\left(1+\frac{\rho_{2z}}{\hat{h}}\right)
\left(
\begin{array}{ccc}
0
&
0
&
\frac{\rho_{1z}}{\hat{h}} - \frac{\rho_{2z}}{\hat{h}}
\\
0
&
0
&
0
\\
\frac{\rho_{1z}}{\hat{h}} - \frac{\rho_{2z}}{\hat{h}}
&
0
&
0
\end{array}
\right),
\label{G12s1}
\qe
and
\eq
\zeta_0 \bG_{12,a}
&\simeq&
\frac{C_D \hat{h}}{2}
\left(1+\frac{\rho_{1z}}{\hat{h}}\right)
\left(1+\frac{\rho_{2z}}{\hat{h}}\right)
\left(
\begin{array}{ccc}
0
&
0
&
-(2 + \frac{\rho_{1z}}{\hat{h}} + \frac{\rho_{2z}}{\hat{h}})
\\
0
&
0
&
0
\\
2 + \frac{\rho_{1z}}{\hat{h}} + \frac{\rho_{2z}}{\hat{h}}
&
0
&
0
\end{array}
\right).
\label{G12a1}
\qe
When we exchange $\phi_1$ and $\phi_2$ or $\brho_1$ and $\brho_2$,
the sign of $\bG_{12,s}$ is reversed,
while $\bG_{12,a}$ remains unchanged to this order.

In the bulk geometry $h/d \to \infty$,
the effects of finite trajectory size can be examined more simply,
by expanding the Oseen tensor to the first order in $\brho$.
It gives
\eq
\zeta_0 \bG(\br_1, \br_2)
&\simeq&
C_{bulk}
\left(
\begin{array}{ccc}
2(1+ \rho_{1x} - \rho_{2x}) &
-(\rho_{1y} - \rho_{2y}) &
-(\rho_{1z} - \rho_{2z})
\\
-(\rho_{1y} - \rho_{2y}) &
1+\rho_{1x} - \rho_{2x}  & 0
\\
-(\rho_{1z} - \rho_{2z}) & 0 &
1+\rho_{1x} - \rho_{2z}
\end{array}
\right)
\label{G12b1}
\qe
with $C_{bulk} = 3a/4d$.
Therefore all the terms change
their signs upon exchanging $\phi_1$ and $\phi_2$.

In the following,
the effective potential is calculated for the circular trajectory
(\ref{circle}) rotated around the $x$-axis by the angle $\beta$, or
\eq
\bR(\phi) = b (\cos \phi, \cos\beta\sin\phi, -\sin\beta \cos\phi).
\qe
The aspect ratio $b/d$ will be fixed to $0.05$ unless otherwise stated.
We use
both the full Blake tensor (\ref{Blake})
and its first-order approximation [(\ref{G12s1}), (\ref{G12a1}), 
and (\ref{G12b1})],
and compare with the zeroth-order (far-field) results.

\subsubsection{Force profile (\ref{Fsin2phi}) with $A_2=0.5$}

\begin{figure}[t]
\includegraphics[width=0.48\textwidth]{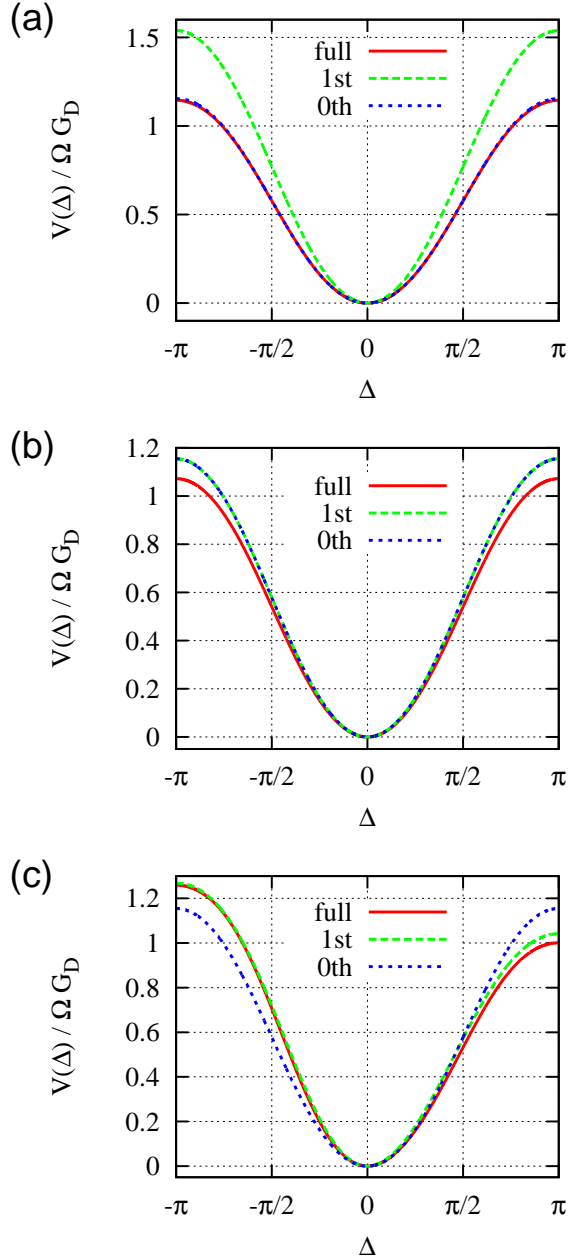}
\caption{
The effective potential $V(\Delta)$
for the circular trajectory (\ref{circle}) with $b/d=0.05$
and the force profile (\ref{Fsin2phi}) with $A_2=0.5$.
The trajectories are either
(a) in the bulk  ($h/d \to \infty$),
(b) horizontal to the substrate ($h/d=0.1$, $\beta=0$),
or
(c) vertical to the substrate ($h/d=0.1$, $\beta=\pi/2$).
Shown are results from the full Blake tensor as well as
its zeroth-order and first-order approximations in terms of $b/d$.
}
\label{fig:A2proximity}
\end{figure}

The effective potential is plotted in Fig. \ref{fig:A2proximity}
for trajectories lying (a) in the bulk ($h/d\to \infty$),
(b) in a plane horizontal to the substrate ($h/d = 0.1$, $\beta=0$),
and (c) vertical to the substrate ($h/d=0.1$, $\beta=\pi/2$).
In (a) and (b), the potential is an even function of $\Delta$
and is well approximated by the zeroth order result (\ref{Vsinphi}),
while in (c), $V(\Delta)$ has a negative average gradient
with $V(\pi) < V(-\pi)$.

\subsubsection{Force profile (\ref{Fsinphi}) with $A_1=0.5$}

\begin{figure}
\includegraphics[width=0.48\textwidth]{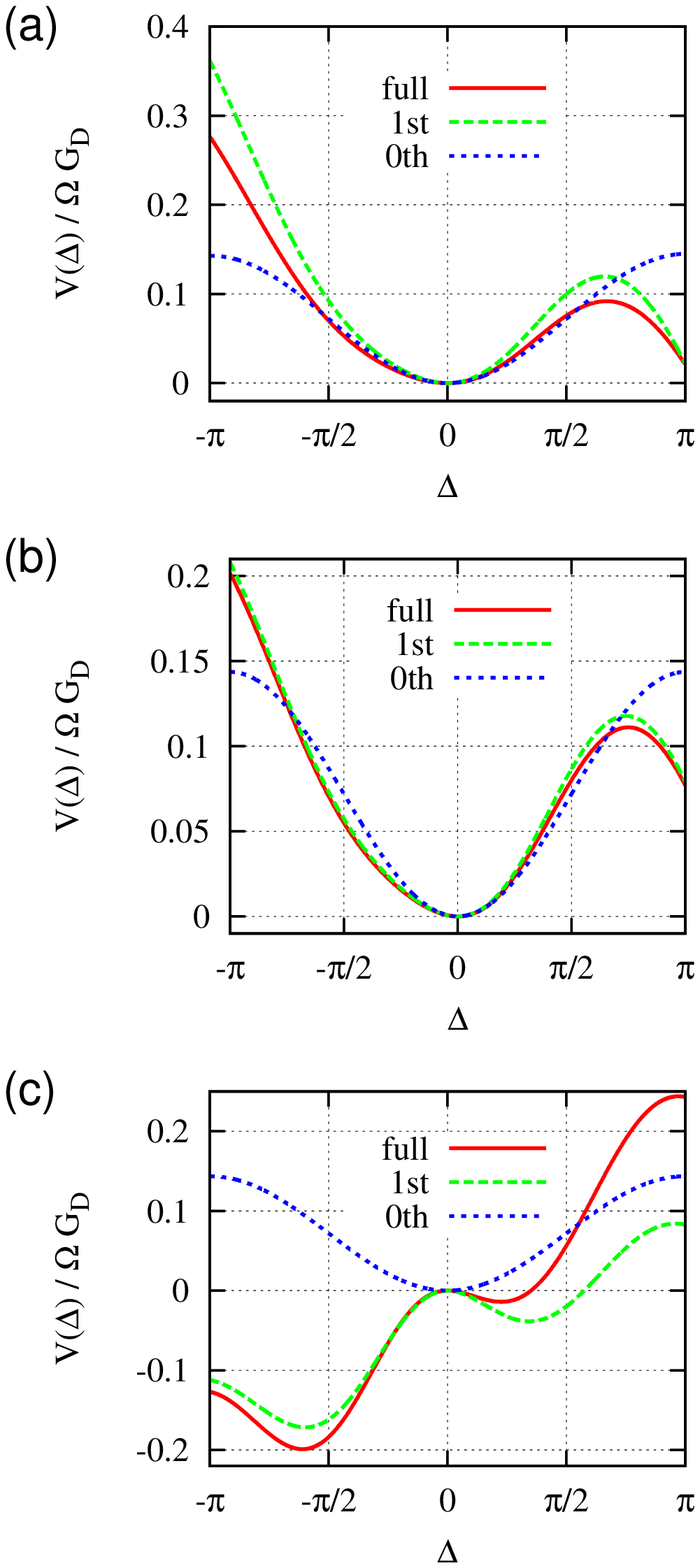}
\caption{
The effective potential $V(\Delta)$
for the circular trajectory (\ref{circle}) with $b/d=0.05$
and the force profile (\ref{Fsinphi}) with $A_1=0.5$.
The trajectories are either
(a) in the bulk  ($h/d \to \infty$),
(b) horizontal to the substrate ($h/d=0.1$, $\beta=0$),
or (c) vertical to the substrate ($h/d=0.1$, $\beta=\pi/2$).
Shown are results from the full Blake tensor as well as
its zeroth-order and first-order approximations in terms of $b/d$.
}
\label{fig:A1proximity}
\end{figure}
\begin{figure}
\includegraphics[width=0.48\textwidth]{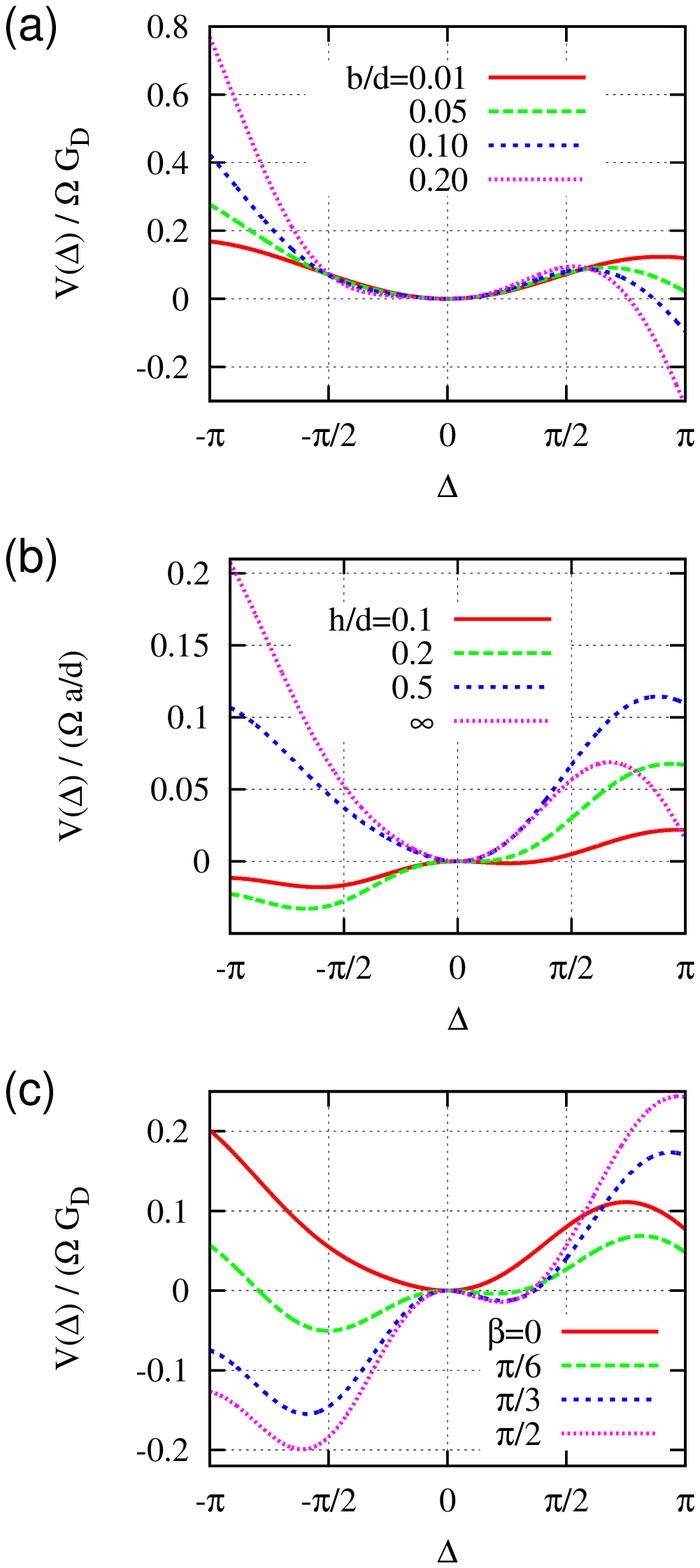}
\caption{
The effective potential $V(\Delta)$
for the circular trajectory (\ref{circle}) with $b/d=0.05$
and the force profile (\ref{Fsinphi}) with $A_1=0.5$.
Shown are dependencies on
(a) $b/d$ in the bulk geometry ($h/d \to \infty$), (b) $h/d$,
and
(c) the tilt angle $\beta$ of the trajectories.
}
\label{fig:A1proximity2}
\end{figure}

The effective potential is plotted in Fig. \ref{fig:A1proximity}.
The trajectories are lying either (a) in the bulk ($h/d\to \infty$),
(b) in a plane horizontal to the substrate
($h/d = 0.1$, $\beta=0$),
or
(c) in a plane vertical to the substrate
($h/d = 0.1, \beta=\pi/2$).
In (a) and (b), the potential curves have negative
average gradient with $V(\pi) < V(-\pi)$, and have
local minima at $\Delta=0$.
Note that the asymmetry of the potential curve
is larger in the bulk case.
In (c), the potential has a positive average gradient
with $V(\pi) > V(-\pi)$, and has two metastable minima.
There is a saddle point at $\Delta=0$.
Note that all these features are already seen in the
first-order approximation.

In Fig. \ref{fig:A1proximity2}(a), we plot the potential
for different trajectory size in the bulk geometry.
The average gradient of the potential is enhanced with $b/d$.
In Fig. \ref{fig:A1proximity2}(b), we plot the potential
for the vertical trajectory ($\beta=\pi/2$) and
with different height.
Note that the average gradient of the potential
changes its sign from positive to negative
at intermediate height.
In Fig. \ref{fig:A1proximity2}(c), we show
the dependence on the tilt angle $\beta$.
The average gradient of the potential
changes its sign around $\beta = \pi/6$.
We thus find that the asymmetry of the potential
sensitively depends on the size, the height,
and the tilt of the trajectories.

\section{Effect of Flexibility}

Hitherto we have only considered rotors with rigid trajectories, 
but in a real system the trajectory could be affected by 
hydrodynamic flow due to flexibility or compliance of the rotor.
As an example, let us consider a bead driven by optical tweezers,
whose focus moves along a prescribed trajectory.
By controlling the distance between the focal point and the bead,
one can tune the tangential driving force~\cite{Bruot}.
We approximate the potential created by the laser beam
by the harmonic potential
$U(\bS) =  \frac{k}{2} S^2$, where $\bS$ is the displacement
of the bead from the focal point.
The bead of the $i$-th rotor is thus positioned at
$
\br_i(t) = \br_{i0} + \bR(\phi_i(t)) +  \bS_i(t),
$
and its velocity is $\dot\br_i = \bR'(\phi_i) \dot\phi_i + \dot{\bS}_i$.
The traction force due to the laser beam is
balanced with the viscous drag force as
\eq
k \bS_i = \bg_i = \zeta_0 [\bv(\br_i) - \dot\br_i],
\label{Ftweezer2}
\qe
while its tangential component is prescribed as $F(\phi_i) = - k \bS_i \cdot \bt(\phi_i)$,
or
\eq
F_i = - \bg_i \cdot \bt_i,
\label{Ftweezer1}
\qe
where abbreviations $F_i = F(\phi_i)$ and $\bt_i = \bt(\phi_i)$ are used as before.
In the limit $k \to \infty$, we restore the model of rigid rotors developed
in the previous sections.
Let us derive the phase evolution equation
by expansion in powers of $k^{-1}$.

First, the intrinsic phase velocity $\omega_i = \omega(\phi_i)$
is determined by
setting $\bv(\br)=0$. Equation (\ref{Ftweezer2}) gives
\eq
\dot{\bS}_i
= -\frac{\zeta_0}{k} \frac{d}{dt}(\bR_i' \omega_i + \dot{\bS}_i)
\simeq -\frac{\zeta_0}{k} \frac{d}{dt}(\bR_i'\omega_i)
\simeq -\frac{1}{k} (F_i \bt_i)' \omega_i
\label{Ftweezer3}
\qe
up to $O(k^{-1})$.
Using this in (\ref{Ftweezer1}),  we obtain
the intrinsic frequency to $O(k^{-1})$ as
\eq
\omega_i
\simeq
\frac{F_i}{\zeta_0 |\bR_i|}
\left(
1 + \frac{F'_i}{k |\bR'_i|}
\right).
\qe
The hydrodynamic interaction is incorporated by
substituting
\eq
\bv(\br_i)
= - \sum_j \bG_{ij} \cdot \bg_j
\simeq - \sum_j \zeta_0 \bG_{ij} \cdot 
\left(\bR'_j \omega_j + \dot{\bS}_j\right)
\qe
into (\ref{Ftweezer2}).
For simplicity
let us assume the far-field limit, where $\bG_{ij}$ is
given by the constant symmetric tensor (\ref{G12FF}).
After some calculation, we obtain the phase evolution equation 
in the form
\eq
\dot\phi_i
&=&
\omega_i \left( 1+ \sum_{j\neq i} J_{ij} \right)
+ \sum_{j \neq i}
\frac{|\bR'_j|}{|\bR'_i|} H_{ij} \omega_j,
\label{Ftweezer4}
\qe
where the function
$H_{ij} = H_{ij}(\phi_i,\phi_j)$ now includes 
an $O(1/k)$ correction as 
\eq
H_{ij} &=& \bt_i \cdot \zeta_0 \bG_{ij} \cdot
\left[ \bt_j - \frac{(F_j \bt_j)'}{k|\bR'_j|} \right],
\qe
and $J_{ij} = J_{ij}(\phi_i,\phi_j)$ is defined by
\eq
J_{ij} &=& \frac{\bt'_i}{k|\bR'_i|} \cdot \zeta_0 \bG_{ij} \cdot F_j \bt_j.
\qe
The stability of the synchronized state is examined
by setting $\phi_1 = \phi + \delta$ and $\phi_2 = \phi$ and
linearizing the evolution equation of $\delta$
as before.
After some straightforward calculation, we obtain
the cycle-averaged growth rate of the phase difference $\delta$ as
\eq
\Gamma =
\frac{1}{T_0}\int_0^{2\pi} d\phi
\left\{
-2 \left[\ln \left(|\bR'| \omega \right) \right]' H_{12}
+ \left(\ln \omega \right)' J_{12}
+ \Delta H'_{12} + \Delta J'_{12}
\right\},
\label{Ftweezer6}
\qe
where the functions in the integrand are
to be evaluated at $\phi_1 = \phi_2 = \phi$
and we define
\eq
\Delta A'
&=&
\left.
\left( 
\frac{\partial A}{\partial \phi_1} 
- 
\frac{\partial A}{\partial \phi_2} 
\right)
\right|_{\phi_1 = \phi_2 = \phi}
\label{Ftweezer5}
\qe
for any two-variable function $A(\phi_1,\phi_2)$.

For example, let us consider the circular trajectory
$\bR(\phi) = b(\cos\phi, \sin\phi, 0) = b \bn(\phi)$.
Using $|\bR'(\phi)| = b$, $\bn'(\phi) = \bt(\phi)$, $\bt'(\phi) = -\bn(\phi)$,
and
the Fourier representation (\ref{Ffourier}) of the force profile,
we obtain the growth rate up to $O(A_n)$ as
\eq
\Gamma =
\frac{2\pi}{T_0}
\left[
A_2 G_D 
\left( \cos \delta_2 - \frac32 \sin \delta_2 \cdot \frac{F_0}{kb} \right)
- (4 G_I + 2 G_D) \frac{F_0}{kb}
\right].
\label{Gammaflexcircle}
\qe
This should be compared to the result
(\ref{Gammacircle}) for the rigid rotors.
We see that the flexibility tends to enhance synchronization
due to the last term on the RHS.
Note also that the small parameter representing the flexibility is $F_0/kb$.
If the displacement from the focal point
(which has the typical magnitude $S_0 \sim F_0/k$)
is much smaller than the size of the trajectory, which
is the case in the optical tweezer experiment~\cite{Bruot},
the flexibility has only a weak effect in inducing synchronization.
These results qualitatively agree with the findings of
the previous study~\cite{NEL08} that assumed constant
driving force and radial displacement from a circular trajectory.
In the paper, the model parameters are estimated for cilia,
which give the dimensionless coupling (that corresponds
to our $F_0/kb$) to be on the order of $10^{-2}-10^{-3}$.
On the other hand, we can expect $O(1)$ modulation of
the driving force from the effective and recovery strokes of cilia.
Therefore, we conjecture that the force modulation plays
dominant roles in establishing the coordinated ciliary beating.
Finally, we mention that our model of flexibility can be
also modified for rotors allowing tangential displacement,
such as a bead attached to the tip of an elastic rod.

\section{Concluding Remarks}

By linear and nonlinear analysis of the coupled oscillator equation,
we have fully characterized the dynamical states of a pair of rotors
making rigid trajectories.
In particular, we obtained the necessary and sufficient conditions for
in-phase synchronization, which show that a wide variety of beating
patterns induce synchronization for an arbitrary trajectory shape.
Even for parallel linear trajectories, which predict only marginal
stability in the linear analysis, the effective potential has a global
minimum at the in-phase state if we choose a suitable force profile.

The results confirm and strengthen our previous finding~\cite{UG11}
that flexibility of the rotors is {\it not} a requisite for synchronization,
although it has been highlighted in many other studies.
In the present paper, we incorporated flexibility into our model
and explicitly compared its effect to the effect of force modulation.
If the disturbance of the trajectory due to hydrodynamic interaction
is small compared to the size of trajectory, the flexibility has
only weak effect in establishing synchronization.
For cilia, sizable modulation of the driving force
is expected from their effective and recovery strokes, and
it should play a dominant role in coordinating their beating.
Recently, another mechanism for driving synchronization
between flagella of a swimming {\it Chlamydomonas} has been 
proposed using a simple three-sphere model \cite{FJ12,BG12}.
In these studies, the phases of the two flagella are predominantly
coupled via translation and rotation of the cell body. 
This type of coupling originates from the condition 
that the net force and torque acting on the cell vanish,
and is specific to rotors attached to a freely-suspended body.
The condition for synchronization with this type of coupling
is different from hydrodynamic synchronization (for example 
even constant forcing profile and circular trajectory 
could lead to synchronization under certain circumstances). 
Also, the coupling is weaker than the hydrodynamic one 
if the cell body is much larger than the distance between
neighboring flagella or cilia, which is the case in 
densely flagellated/ciliated cells such as {\it Volvox} 
and {\it Paramecium}.

The effective potential that governs the nonlinear
dynamics of the rotors have a number of remarkable features.
First, it allows us to locate all the stable and metastable
states of the system at a glance.
In the far-field limit, the potential is symmetric.
For circular trajectories with simple force modulation
(consisting of a single harmonic mode),
the potential has only one minimum that describes
either in-phase or anti-phase synchronization.
Bistable and metastable states appear for more complex
trajectory shapes such as ellipses.
When the system is trapped in an out-of-phase
stable/metastable state, the phase difference
(in a natural gauge) oscillates as a function of time.
We have incorporated near-field corrections
due to finite trajectory size,
and found that the overall shape of the potential,
especially its average gradient, sensitively
depends on the size/height/tilt of the trajectory.
When the potential has a non-vanishing average gradient,
each of its local minimum corresponds to a metastable state.
In the presence of strong noise, we may observe phase slippage
in a specific direction. We note that experiments on the flagellar beating
of a mutant {\it Chlamydomonas} have recently shown anti-phase
synchronization \cite{Goldstein2}. It will be interesting to probe
the differences between this mutant strain and the wild-type
{\it Chlamydomonas} in terms of the beating pattern of the flagella,
and examine whether the phenomenon can be quantified within the framework
of our model.

A more direct experimental test of our findings could be pursued
in a simpler system that does not have the complexity of
the living organisms, such as optically driven colloids.
Optical tweezers with moving focus can drive the colloidal particle
on a prescribed trajectory, and by controlling the distance between
the focal point and the bead, one can also prescribe the force profile.
Experiments are currently underway in the group of Pietro Cicuta
at the Cavendish Laboratory along these lines~\cite{Bruot}.
Also, optical vortices~\cite{Gahagan96,Curtis03} have been used to trap
colloidal particles on a ring and drive them in one direction.
The driving torque could be modulated by tailoring the helical
structure of the laser beam to give a prescribed force profile.

In forthcoming papers, we plan to discuss the collective dynamics
of arrayed rotors, and in particular the formation of traveling waves.
Such a study should become a first step towards understanding
the relation between the beating pattern of cilia
and the metachronal waves they form.
We will also consider a pair of rotors
with different intrinsic frequencies,
which will induce phase slips similar to those
observed in {\it Chlamydomonas}~\cite{Goldstein-Science-09, GP09}.
The present paper assumes spherical beads,
but the analysis could be extended to
non-spherical bodies such as rods or helices,
which are closer to the shapes of biological filaments.

\acknowledgments
NU acknowledges the support by JSPS KAKENHI (Grant Number 23740286)
and the JSPS Core-to-Core Program ``International research
network for non-equilibrium dynamics of soft matter''.



\begin{thebibliography}{99}
\bibitem{note_linear_trajectory}
By considering nonlinear effect, we show in Section IV
that linear trajectories are also capable of synchronization.
\end{thebibliography}
\end{document}